\documentclass[final,3p,times,twocolumn]{elsarticle}

\usepackage{amssymb}
\usepackage{amsmath}

\journal{Nucl. Instrum. and Meth. in Phys. Research A}

\begin{document}

\begin{frontmatter}



\title{Performance of a prototype bakelite RPC at GIF++ using self-triggered electronics for the
CBM Experiment at FAIR}
\author[label1,label2]{Mitali Mondal}\corref{cor1}
\cortext[cor1]{Corresponding author}
\ead{mitalimondal.phy@gmail.com}
\author[label1,label2]{Tanay Dey}
\author[label1]{Subhasis Chattopadhyay} 
\author[label1]{Jogender Saini}
\author[label1]{Zubayer Ahammed}
\address[label1]{Variable Energy Cyclotron Centre, 1/AF, Bidhannagar, Kolkata - 700064, India}
\address[label2]{Homi Bhabha National Institute, Training School Complex,
Anushakti Nagar, Mumbai 400094, India}

\begin{abstract}
The Muon Chamber (MuCh) is a sub-system of the Compressed Baryonic Matter (CBM) experiment for the
detection of low-mass-vector mesons produced in high energy heavy ion collisions at beam energies
ranging from 2 AGeV to 11 AGeV and decaying in the di-muon channel. MuCh consists of a segmented 
absorber and four detector triplet stations sandwiched between the absorber segments. At the 
3$^{rd}$ and 4$^{th}$ stations of MuCh, Resistive Plate Chambers
(RPCs) have been conceived for muon tracking. We have tested the performance of a low resistivity 
bakelite RPC prototype equipped with self-triggered front end electronics (MuCh-XYTER) for the CBM
Muon Chamber. A systematic study on the muon detection efficiency and time resolution has been 
carried out in a high-rate photon background at the Gamma Irradiation Facility (GIF++) at CERN. 
The details of the measurement setup and the results are presented here.
\end{abstract}

\end{frontmatter}
\section{Introduction}
\label{Intro}
Compressed Baryonic Matter (CBM) ~\cite{a,cbm1} is set to be a fixed target experiment at the 
Facility for Anti-proton and Ion Research (FAIR) centre currently under construction in Darmstadt, Germany. The 
experiment will be dedicated to the study of nuclear matter in the region of high net baryon densities
and moderate temperatures likely to be created at FAIR by collisions of heavy ion beams with 
targets of high mass number with beam energies ranging from 2 AGeV to 11 AGeV~\cite{b}. The uniqueness
of the CBM experiment is the aim to handle a very high interaction rate up to 10 MHz, which
exceeds the well-known high energy nuclear collision experiments carried out at STAR, ALICE, NA61
by several orders of magnitude~\cite{highrate}. This facilitates high statistics measurements of the
rare diagnostic probes of the collision medium~\cite{a}. To cope with this high interaction rate, 
fast and radiation hard detectors, high-speed trigger and data acquisition systems and novel 
real-time event reconstruction techniques are being developed~\cite{c}. Muon Chamber (MuCh) is an
important part of the CBM detector system for tracking and identifying muon pairs formed via the
decay of produced particles like charmonium (J/$\psi$) and low mass vector mesons (LMVM)~\cite{d}.
The rate of incident particles per unit area on the detector surface decreases radially and
as we go away from the target in Z direction. For the 1$^{st}$ and 2$^{nd}$ MuCh stations, Gas
Electron Multiplier (GEM) chambers will be used~\cite{cbmgem}. Resistive Plate Chambers (RPCs)~\cite{rpc}
have been proposed as a technology option for the 3$^{rd}$ and 4$^{th}$
stations of MuCh. Simulation results indicate the hit rates on the 3$^{rd}$ and 4$^{th}$ stations of MuCh reach
a maximum of 13 kHz$/$cm$^2$ and 4.7 kHz$/$cm$^2$ respectively for
minimum-bias Au-Au collisions at a beam energy of 10 AGeV. The RPCs in CBM, therefore, should be capable
of handling these high rates for muon detection with high efficiency ($\sim$ 90 $\%$). Cosmic ray experiments like Cover
Plastex~\cite{pla} in UK, Pierre Auger Cosmic Ray Observatory~\cite{pierre} in Argentina, neutrino physics
experiments like Daya Bay Reactor in China~\cite{daya} and INO-ICAL in India~\cite{ino}, high energy physics
experiments like CMS, ALICE, ATLAS at LHC~\cite{lhc}, BaBar at PEPII~\cite{pep} and BELLE at KEKB~\cite{kekb}
are using or planning to use RPCs for muon detection at various particle rates. These detectors are 
preferred for muon detection due to several reasons,\newline
\hspace*{0.8 cm}(a) good time resolution,\newline
\hspace*{0.8 cm}(b) fast response time,\newline
\hspace*{0.8 cm}(c) high detection efficiency,\newline
\hspace*{0.8 cm}(d) simple and less expensive construction compared to other gaseous detectors like GEM or Micromegas, and\newline
\hspace*{0.8 cm}(e) the possibility to build them in large areas and in different shapes~\cite{rpc}.\newline 

The inherent time constant of RPCs, i.e., the time duration for which the detector region remains blind is very high due to the high 
resistivity (~10$^{12}$ $\Omega$ cm ) of the standard soda-lime glass or bakelite electrodes~\cite{BESIII,resis,ying}. This does not allow
it to be used at high particle rate. This paper explores the possibility of using RPCs at high particle flux (upto $\sim$ 13 kHz$/$cm$^2$)
for the CBM experiment at FAIR. We have tested a relatively low bulk resistivity ($\sim$ 10$^{10}$ $\Omega$ cm) single-gap bakelite RPC
module in a beam test at GIF++ at CERN using self-triggered MuCh-XYTER~\cite{f} electronics. This particular readout ASIC is highly prone
to noise and therefore requires a low-noise detector system for optimum performance.

\section{Rate capability for a RPC}
\label{RateRPC}
The rate capability for a RPC ($\Re$) is limited by the drop in voltage ($V_d$) across the resistive electrodes.
This voltage drop is determined by the simplistic static model (Ohmic model),
\begin{equation}\label{eq1}
\begin{split}
V_d &= V_a - V_{gap}\\
& = iR\\
& = \Phi\,\rho\,t\,\langle Q \rangle  
\end{split}
\end{equation}
where V$_a$ and V$_{gap}$ are the voltages applied across the chamber and to the gas gap
respectively. A current $\it i $ produces the voltage drop V$_d$ across the electrodes of total resistance R.
Thus the working current $\it i $  driven by the highly resistive electrode plates limits the rate capability of RPC.
To increase the rate capability under heavy irradiation (particle flux $\Phi$), V$_d$ needs to be
kept at a minimum value with respect to V$_{gap}$, which can be done by reducing any combination of the three
detector parameters, the bulk resistivity of the electrode material ($\rho$), the total thickness of both the
electrodes ($\it t $) and the average charge produced in the gas for each incident particle($\langle$ Q
$\rangle$)~\cite{aielli,preamp}.

We have taken the approach towards the reduction of average charge $\langle$ Q $\rangle$~\cite{avQ} and
the reduction in the bulk resistivity of electrodes to achieve the required rate capability. In this context
it should be mentioned that a reduction of the gap size decreases the charge content ($\langle$ Q $\rangle$)
as well and could be also used to increase the rate capability. A sensitive front-end electronics, which can detect
low amount of charge, needs to be used to operate an RPC in a low gain mode~\cite{h} in order to maintain the anticipated
efficiency. In this paper, a low gain operation of a low bulk resistivity
($\sim$ 10$^{10}$ $\Omega$ cm) bakelite RPC along with a highly sensitive front-end electronics has been pursued
to operate the RPC in a high rate environment.   

\section{GIF++ test beam setup}
\label{GIF++}
\begin{figure}[htbp]
\centering
\includegraphics*[height=70mm,width=78mm, keepaspectratio]{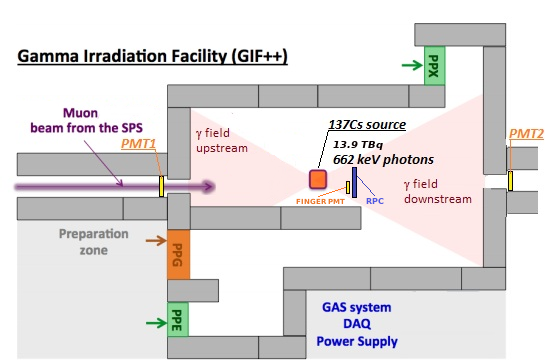} 
\caption{\label{gif++}[Colour online] Block Diagram of the GIF++ setup.}
\end{figure}

The Gamma Irradiation Facility (GIF++) is located at the H4 beamline in the Super
Proton Synchrotron (SPS), North area, at CERN. A schematic block diagram of the GIF++ setup is 
shown in Fig. \ref{gif++}. It consists of an intense $^{137}$Cs gamma source of activity 13.9 TBq 
with 
different attenuation filters to limit the gamma flux. 662 keV photons are generated from the
source with 85$\%$ branching ratio. Three layers of filters have been used in front of the source
and each layer contains three slots of absorbers (not shown in the figure). The absorbers are made of
Al or Pb. GIF++ also provides a high momentum muon beam (100 GeV/c) to test the performance of
detectors for high-energy physics experiments in the presence of high rate background.

\begin{figure}[htbp]
\centering
\includegraphics*[height=80mm,width=80mm, keepaspectratio]{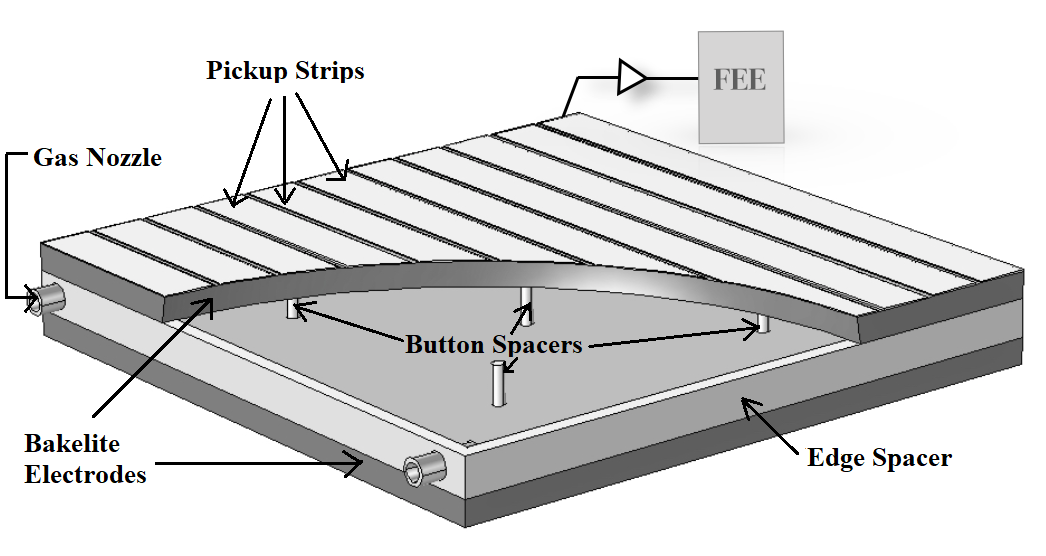} 
\caption{\label{schematic-rpc}[Colour online] A schematic diagram of the chamber. The electrodes are
separated by spacers to form a gas-gap of 2 mm. The chamber is read out using a pick-up panel of 11
strips, each of 2.3 cm width.}
\end{figure}

In this test beam, a $\sim$(30 cm $\times$ 30 cm $\times$ 0.2 cm) single gap RPC
with electrodes made of double-layer linseed oil coated, high-pressure laminate (HPL)
phenolic resin (bakelite), built by the Italian company General Tecnica has been used. Each electrode
thickness was $\sim$ 2 mm and gas-gap thickness was $\sim$ 2 mm. A schematic view of the
chamber with strip-readout is shown in Fig. \ref{schematic-rpc}.

\begin{figure}[htbp]
\centering
\includegraphics*[height=80mm,width=78mm, keepaspectratio]{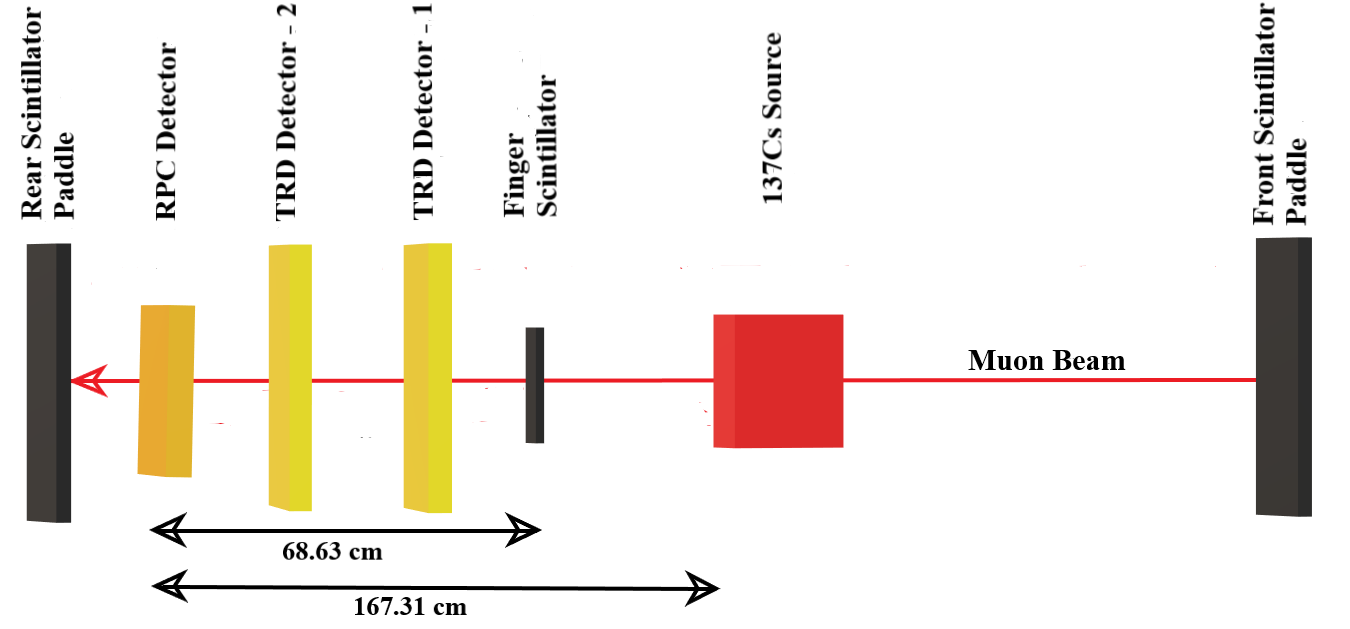} 
\caption{\label{expt-setup}[Colour online] The schematic experimental Setup at GIF++ with RPC, scintillators, gamma source
and Transition Radiation Detectors (TRD).(Not to scale)}
\end{figure}

The bulk resistivity of the electrodes ($\sim$ 4 $\times$ 10$^{10}$ $\Omega$ cm) at room temperature is lower
by 2 orders of magnitude compared to that of a low-rate single-gap bakelite RPC such as those used in cosmic ray
experiments. The detector signals were read out from the anode side only using 11 copper pickup strips. Each strip
width was 2.3 cm, and strip pitch was 2.5 cm. The pickup panel (30 cm $\times$ 30 cm) was made of a 1.5 mm thick FR4
sheet sandwiched between two 35 $\mu$m
copper layers. The chamber was operated in avalanche mode with a gas mixture of R134a: iC$_4$H$_{10}$ : SF$_6$
:: 95.2$\%$ : 4.5$\%$ : 0.3$\%$. To maintain a constant bulk resistivity of the HPL plates, water vapour was
added to the mixture in order to sustain a relative humidity of 60$\%$. This particular gas mixture was
shared by the CMS-RPC which was being tested in the same setup. Our detector prototype was placed vertically
on an aluminium structure on the muon beamline at a distance of $\sim$ 1.67 m from the source position, as
shown in Fig. \ref{expt-setup}. To control the noise due to electromagnetic interference (EMI), the detector
was placed inside a copper box. The copper box was then inserted into a perspex box for mechanical stability
and environmental control. The readout electronics board was attached to the outer surface of the perspex box.

For online triggering, a coincident signal from three scintillators, two paddle scintillators outside the
experimental hall and one finger scintillator of dimesion 11 cm $\times$ 3 cm inside the hall, was used. The positions of the scintillators had been
optimised in such a way that they had an overlap with the area of the detector exposed to the muon beam.

When only the muon beam was present, a CAEN module N471A with a current resolution of 1 nA was used to ramp up
the RPC as this particular module was suitable to handle very low chamber current of the order of 2 $\mu$A at
10.1 kV. During operation in presence of high photon flux along with the muon beam, another CAEN module
N1470A, with a current resolution of 50 nA and a current limit of 3 mA, was used. The chamber was tested at
several photon fluxes, i.e. 18 kHz/cm$^{2}$, 40 kHz/cm$^{2}$, 59 kHz/cm$^{2}$, 87 kHz/cm$^{2}$, 
128 kHz/cm$^{2}$, 187 kHz/cm$^{2}$, 275 kHz/cm$^{2}$ and 403 kHz/cm$^{2}$. The estimation of photon fluxes was
based on the interpolation of known simulated fluxes for the GIF++ facility at various distances from the
source.

\section{Readout chain at GIF++}
\label{DAQ}
\begin{figure}[htbp]
\centering
\includegraphics*[height=90mm,width=80mm, keepaspectratio]{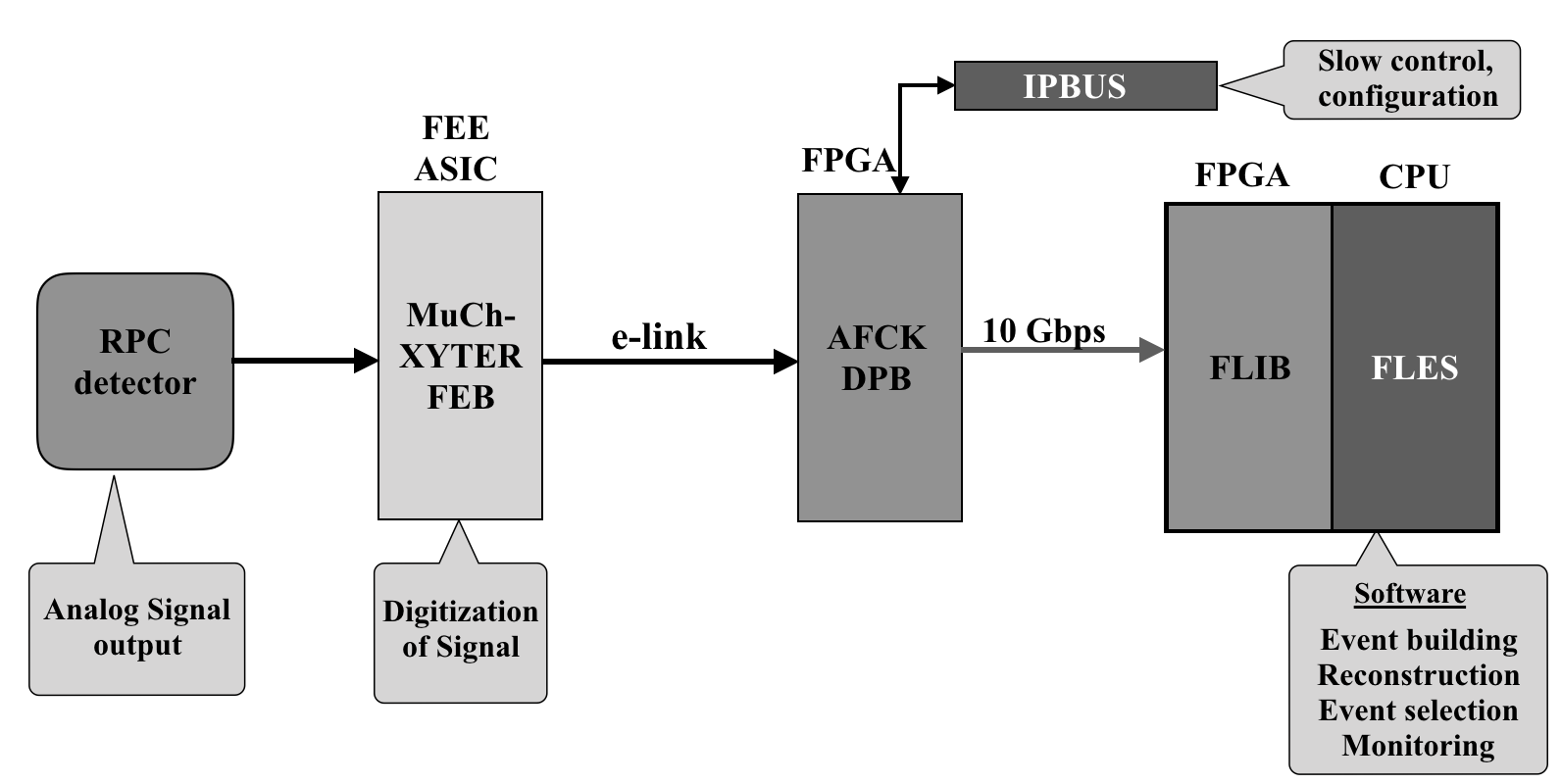} 
\caption{\label{schematic_daq}[Colour online]Data acquisition chain of the prototype RPC at GIF++.}
\end{figure}

A block diagram of the readout chain of the tested RPC as planned to be installed in the CBM experiment is shown in Fig.
\ref{schematic_daq}. It consists of three main blocks. These are$\colon$ the Front End Board (FEB), connected
directly to the chamber and consisting of a self-triggered ASIC called MuCh-XYTER~\cite{asic} for processing
and digitization of the RPC signals; an FPGA-based data processing board called AFCK-DPB; the First Level
Event Selector (FLES), with its interface board FLIB~\cite{afck}.  

\begin{figure}[htbp]
\centering
\includegraphics*[height=60mm,width=78mm, keepaspectratio]{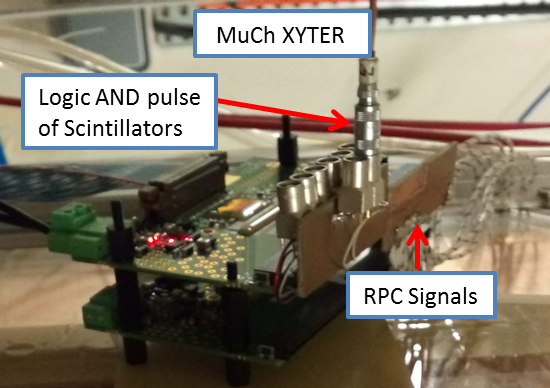} 
\caption{\label{FEE_pic}[Colour online] Front End Electronics of the RPC for the GIF++ beam test.
The FEB is shown connected to the RPC.}
\end{figure}

The self-triggered ASIC in the FEB shown in Fig. \ref{FEE_pic}, called MuCh-XYTER, does not need any external trigger
and collects all detector signals above a predefined threshold. This particular ASIC will also be used for
reading out the GEM chambers in the 1$^{st}$ and 2$^{nd}$ stations of CBM-MuCh. The use of a common chip
in the entire CBM-MuCh avoids the non-uniformity and complexity involved in using different readout systems.
This, however adds a challenge to the operation of RPCs, as these will have to be operated at low gain in
order to match the available dynamic range of the ASIC. Some of the main features~\cite{j,kas16} of the ASIC are listed
below.\newline
(i) Maximum acceptable data of 10 million samples per second for 128 channels with 1 e-link,\newline
(ii) low power consumption ( 10 mW/ch),\newline
(iii) low-noise Charge Sensitive Amplifier (CSA) (nominal charge gain 1.6 mV/fC),\newline
(iv) a time resolution of 3.125/$\sqrt12$ ns,\newline
(v) dynamic range up to 100 fC and,\newline
(vi) radiation-hardened design.

Each ASIC consists of 128 analog channels which can convert analog signals into their corresponding digital
forms. Each channel of the ASIC consists of a charge sensitive amplifier (CSA), a polarity selection circuit
and two parallel signal processing paths defined by two pulse shaping amplifiers. The first path is dedicated
to the timing measurement with a fast shaper and a discriminator. The ASIC provides digitized signals with a 14-bit timestamp.
The second path measures the amplitude of the input pulse using a slow shaper and a 5-bit flash ADC~\cite{k}. 
The time constant of the fast and slow shapers were set to 30 ns and 280 ns respectively. 
We have put a threshold of 50 fC on the ASIC and calibrated it up to 150 fC for this study.
A total of 12 channels from the available 128 channels of the ASIC were used in the test beam. Out of those
12 channels, 11 channels were used for RPC signals and 1 channel for the scintillator trigger. We would like to
point out that the timestamp of the trigger signal is used as the reference for the event. The signal from the
coincidence of scintillators is processed such that it can be used as input to the ASIC according to the available
dynamic range.

As mentioned, the self-triggered electronics will
process any signal that crosses the threshold and transmit it to the data stream. In this scenario, if the threshold crossing
is too frequent, this will consume the usable bandwidth. For the present study, we used a threshold of 50 fC to
reduce the data rate within an acceptable limit. 

\section{Beam test results}

In this section, the results for the detection of muons with and without the presence of a photon
flux at GIF++ are presented. In sub-section 5.1, the voltage-dependent dark current, as obtained
in absence of beam or photons, is discussed. In sub-section 5.2, the time resolution and efficiency
of the detector for detection of muons, in absence of photons, are studied. In sub-section 5.3,
the cluster size is discussed. In sub-section 5.4, the effect of photon fluxes on efficiency is explored
in detail.

\subsection{Current}
Dark current is the current measured by the HV power supply after applying bias in the absence of muon
beam or photon source. The dark current of the RPC module was
\begin{figure}[htbp]
\centering
\includegraphics*[height=60mm,width=78mm, keepaspectratio]{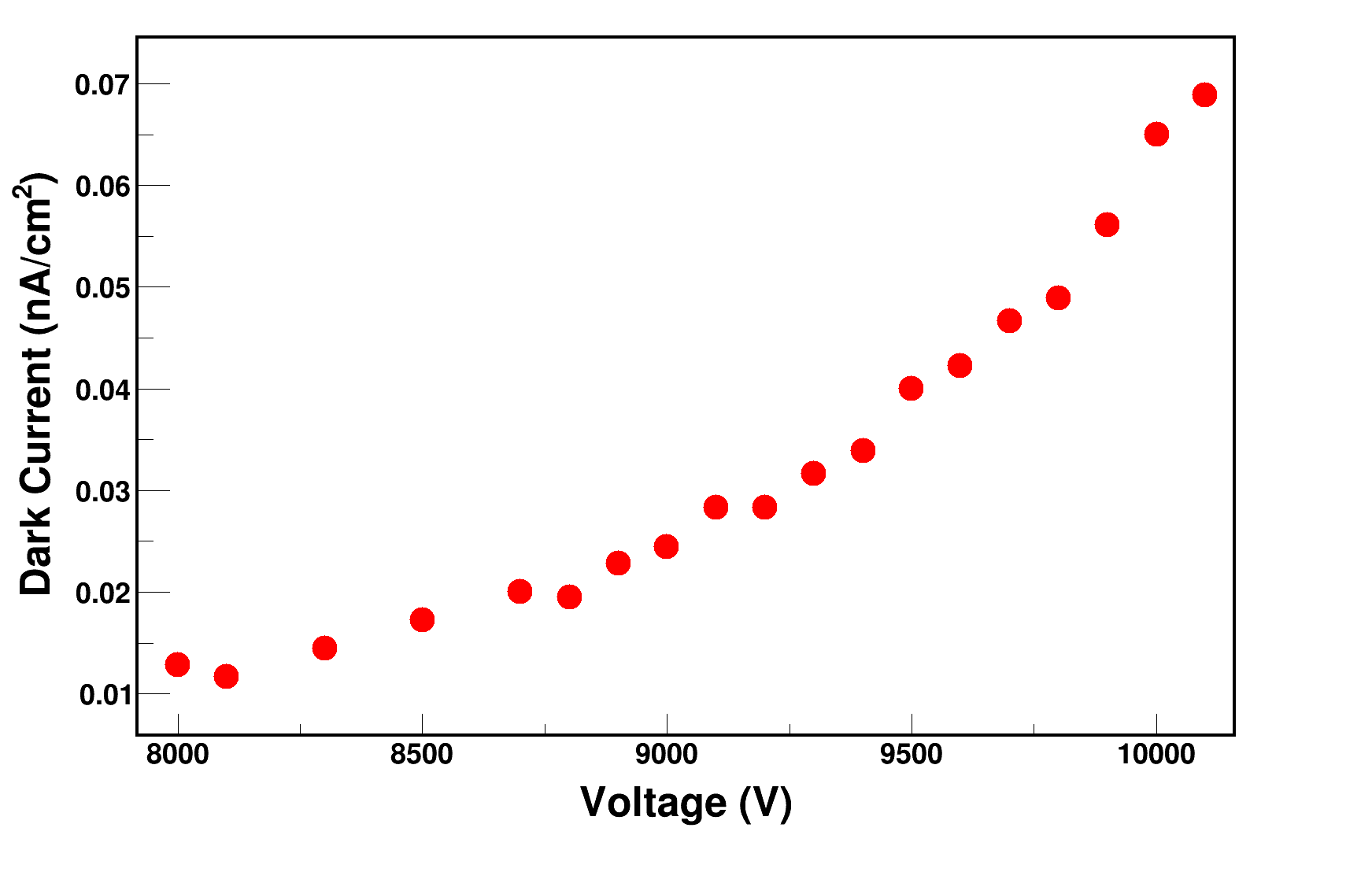} 
\caption{\label{dark}[Colour online] Dark Current per surface unit of the RPC, measured as a function of HV.}
\end{figure}
monitored over the entire HV range. The dark current per surface unit as a function of HV is shown in Fig. \ref{dark}.
At 10 kV, the dark current of the detector per surface unit is 0.07 nA/cm$^2$. The threshold setting of 50 fC brings down the
noise rate to 0.5 Mbps which corresponds to 77 Hz/cm$^2$ at 10 kV. The ALICE single-gap RPCs tested with a highly saturated
avalanche mode gas mixture show an average dark current of 0.05-0.07 nA/cm$^2$~\cite{alicerpcdarkcurrent,alicerpcdimension}, which is similar to that found in our study. 
The maximum detector currents per surface unit have been found to be 2.2 nA/cm$^2$ at 10.1 kV and 
16.7 nA/cm$^2$ at 9.9 kV in presence of the muon flux and photon flux($\sim$ 403 kHz/cm$^2$) respectively.

\subsection{RPC time resolution and efficiency with muon beam}

In this section we discuss the time-resolution and efficiency of the detector as tested with a muon
beam of 10$^5$ particles per spill of 4.5 s duration. The muon flux as obtained from the
scintillator count and the beam profile is about 60 Hz/cm$^2$.

\begin{figure}[htbp]
\centering
\includegraphics*[height=63mm,width=78mm, keepaspectratio]{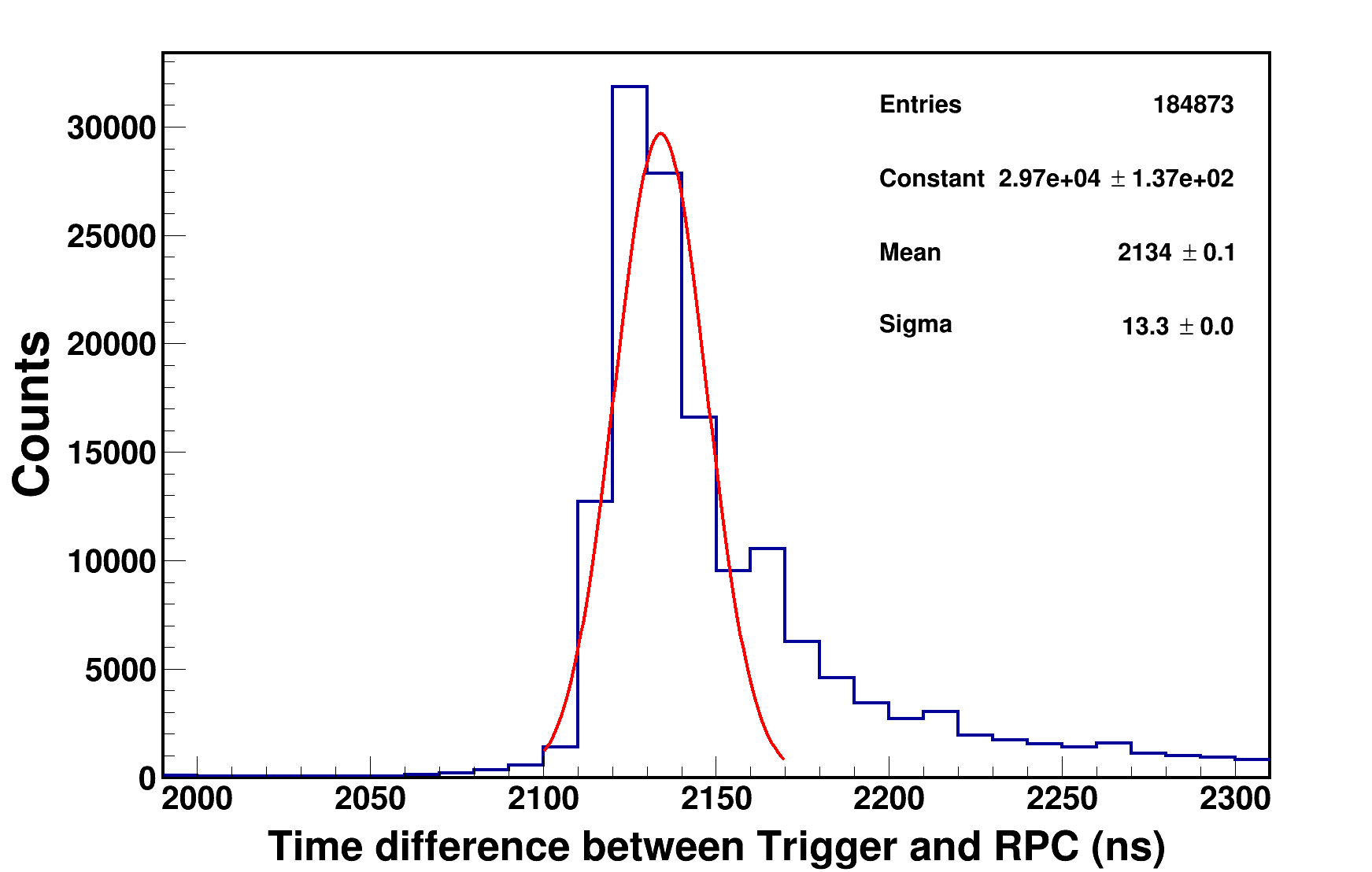} 
\caption{\label{time_diff} [Colour online] Time difference spectra between the RPC detector at
9200V and the trigger. The blue histogram is the measured data and the red one is a Gaussian fit to the data.}
\end{figure}
The analysis was carried out in the CBMROOT environment. In the offline analysis, the time differences between the
RPC pulses and the trigger pulses were obtained. Fig. \ref{time_diff} shows the time difference spectra
between the measured RPC signals and the trigger signals. The tail of the spectra is due to the combination of
late events and streamer contamination. The peak was fitted with a Gaussian distribution to obtain the peak
position and standard deviation ($\sigma$). The peak position represents the delay introduced between the RPC
pulse and the trigger pulse. The fitted $\sigma$ of the time-difference distribution represents the time
resolution of the RPC folded with that of the scintillator trigger. In our study, at an HV of 9.2 kV, we
obtain $\sigma$ of $\sim$ 13 ns. The time resolution, even after unfolding the effect of the
scintillator trigger, appears higher than that of a standard single gap ($\sim$ 2 mm) RPC. Possible reasons for such a
result are (a) operation of the chamber at relatively low gain resulting in higher time resolution, (b) use of
a CSA in the MuCh-XYTER with relatively higher time constant, leading to a bigger time-walk, (c) due to the high
system noise, a threshold of 50 fC has been used  to keep the bandwidth under control. Such a high threshold leads
to relatively large time-walk effects. The measured time resolution is still acceptable for CBM, since the GEM chambers
of the 1$^{st}$ two MuCh stations have a similar time resolution~\cite{gemreso}.

For the efficiency computation, an RPC hit was considered detected if the time difference lies within $\pm$ 25 ns
of the peak, which corresponds to about 2$\sigma$ in the distribution of Fig. \ref{time_diff}. The efficiency
($\eta$) has been calculated as 
\begin{equation}
\eta = \frac{RPC\;hit\;count\;within\;time\;window}{Trigger\;count}
\end{equation}
\begin{figure}[htbp]
\centering
\includegraphics*[height=60mm,width=78mm, keepaspectratio]{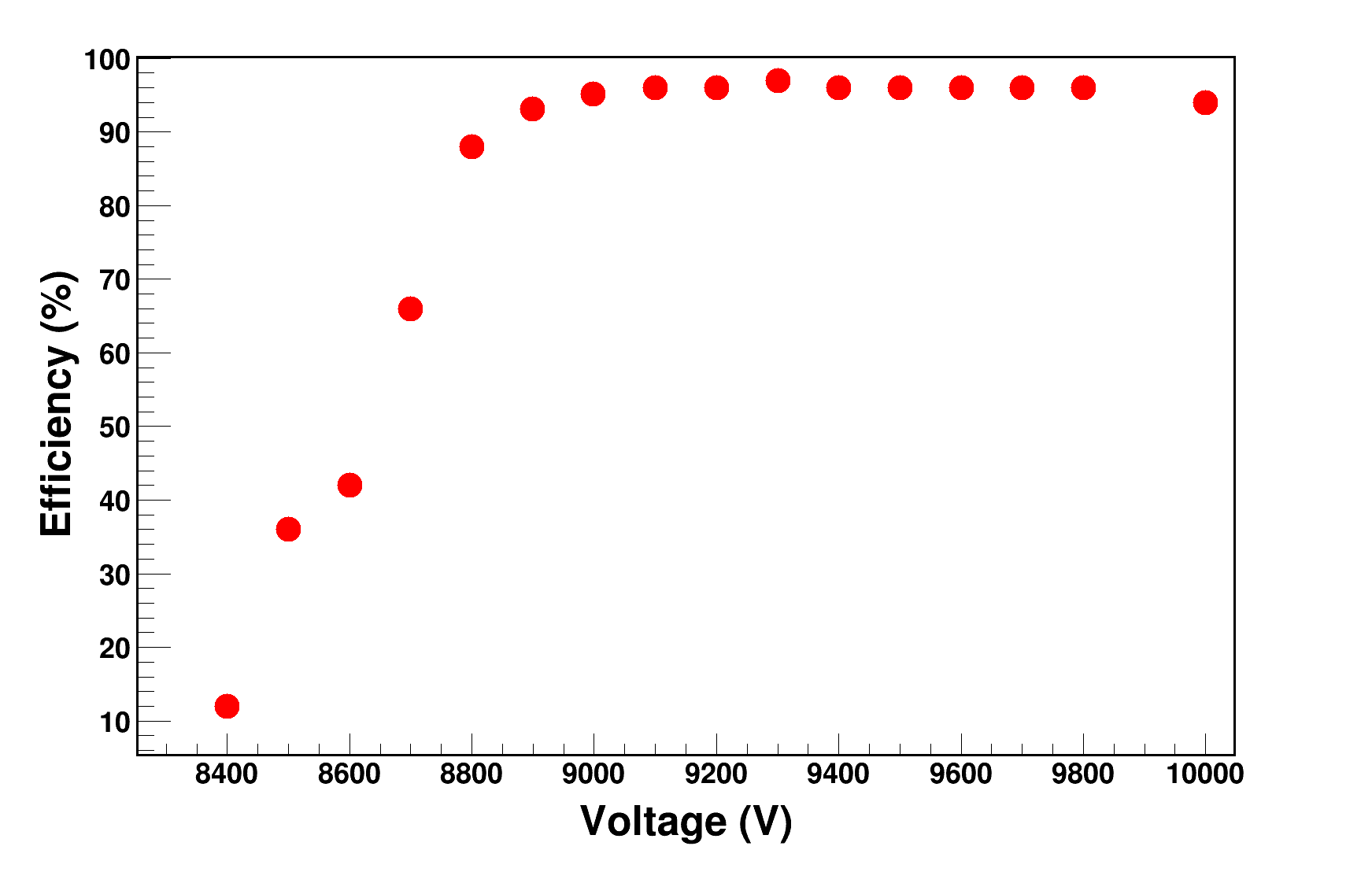}
\caption{\label{efficiency}[Colour online] Muon detection efficiency of the RPC as a function of
the high voltage when exposed to the muon beam alone.}
\end{figure}
Fig. \ref{efficiency} shows the muon detection efficiency of the prototype RPC as a function of
high voltage. The efficiency reached 95$\%$ at 9 kV. The efficiency curve presents a plateau
between 9 kV and 10 kV, which is sufficient for the safe operation of the detector. The efficiency
plateau starts at 9 kV, which is substantially lower when compared with similar RPCs readout
by other FEE, ATLAS RPCs for instance operate at 9.6 kV with 96$\%$ efficiency using a
 similar gas mixture ratio R134a: iC$_4$H$_{10}$ : SF$_6$ :: 95.0$\%$ : 4.7$\%$ : 0.3$\%$~\cite{atlas}.
We are successful in operating the detector at a lower gain compared to other RPC setups.

\subsection{Cluster size}
\begin{figure}[htbp]
\centering
\includegraphics*[height=54mm,width=75mm]{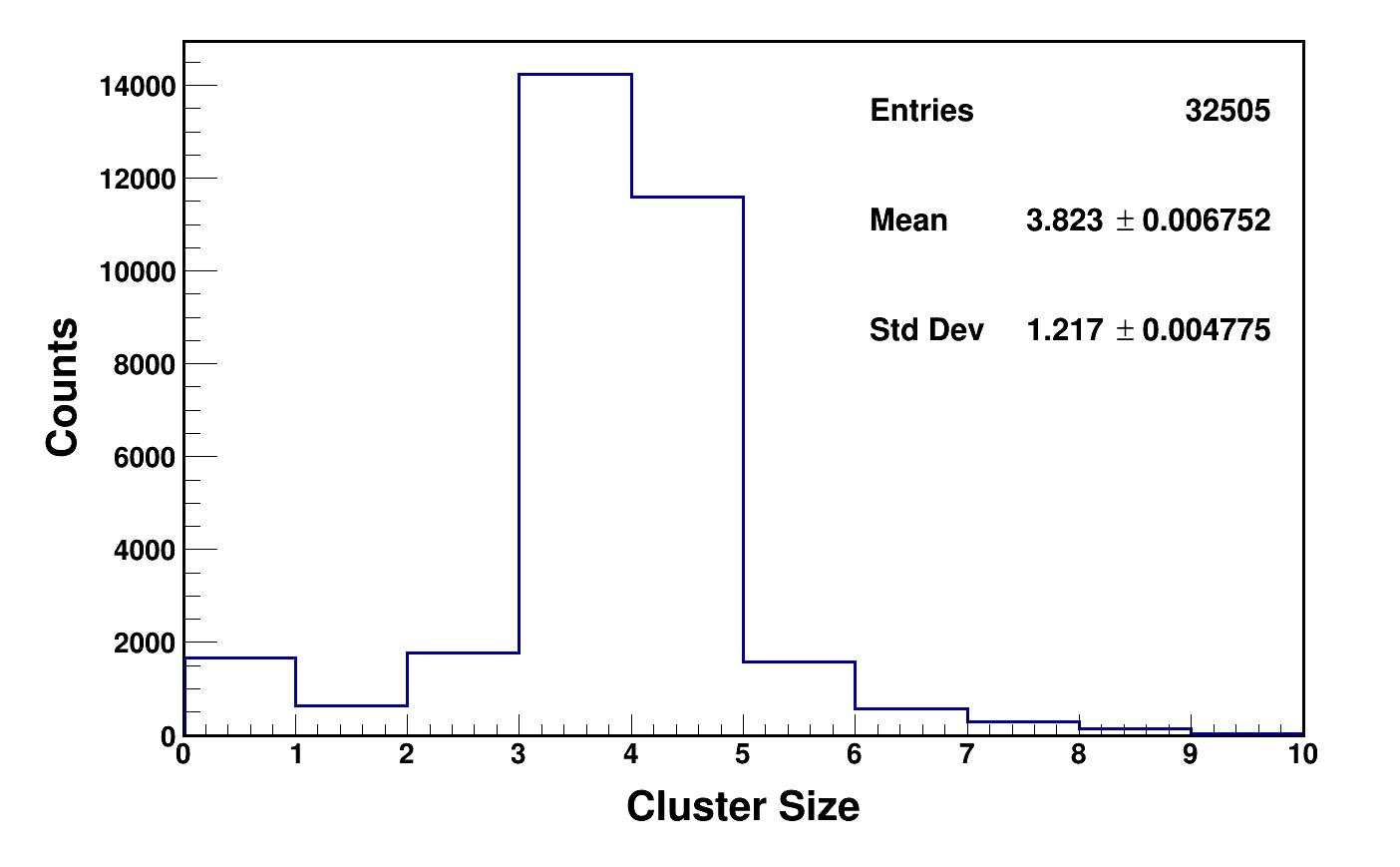}
\caption{\label{clstr_size}[Colour online] Cluster size of muons for RPC at 9200 V}
\end{figure}
The cluster size is defined as the number of RPC strips hit in a single event. The measured average
value for strips of 2.3 cm is around 3.8 for 96$\%$ efficiency. Fig. \ref{clstr_size} shows the distribution
of the cluster size of signals correlated with a  trigger signal at 9.2 kV in the presence of muon beam. 

We conclude from the tracking simulation in CBM with MuCh that the measured cluster size is compatible with the tracking 
requirements of CBM.

\subsection{Rate capability}
\begin{figure}[htbp]
\centering
\includegraphics*[height=60mm,width=78mm, keepaspectratio]{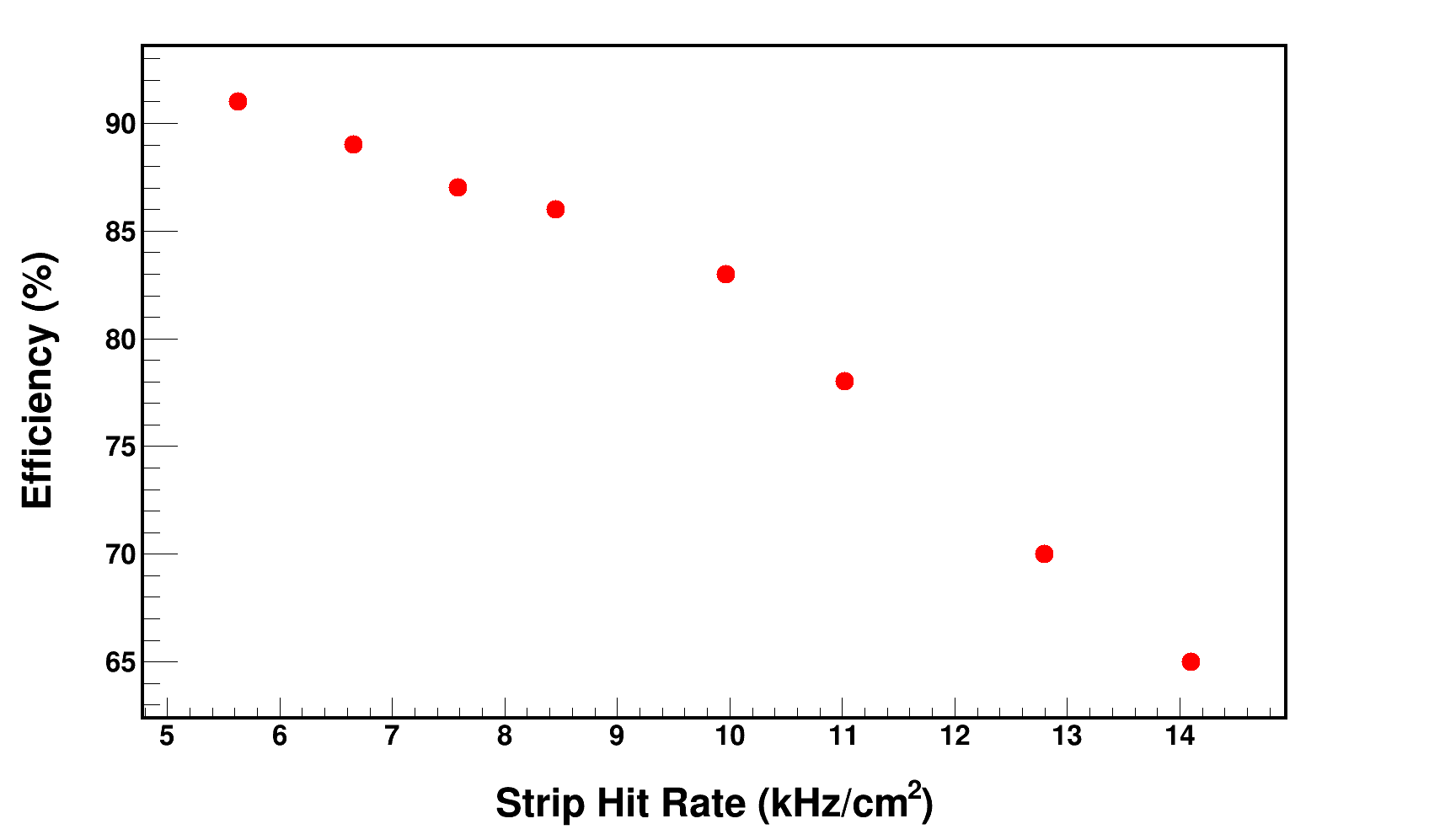} 
\caption{\label{hit_rate}[Colour online] Muon detection efficiency as a function of the strip hit rate at 9.5 kV.}
\end{figure}

The RPC Rate capability has been studied by measuring the detection efficiency of muons as a
function of the photon flux. We have calculated the flux of photons impinging on the detector mounted
at a distance of 1.67 m from the $^{137}$Cs source of activity 13.9 TBq to be about 40 MHz/cm$^{2}$.
Attenuators were placed to reduce the flux on the detector.
\begin{figure}[htbp]
\centering
\includegraphics*[height=75mm,width=80mm, keepaspectratio]{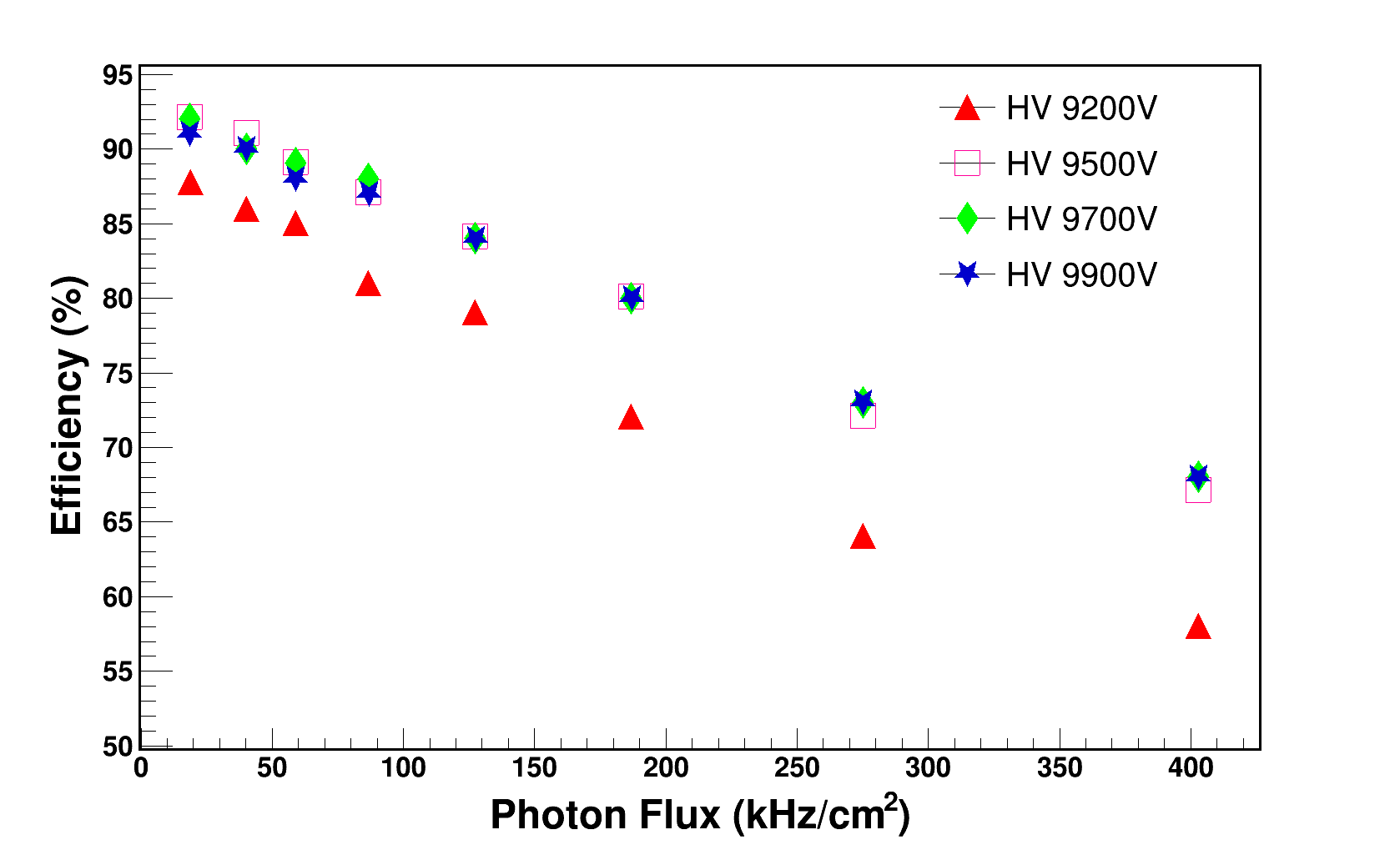} 
\caption{\label{eff_vs_flux}[Colour online] Muon detection efficiency as a function of the photon flux for a
range of high voltages.}
\end{figure}
Fig. \ref{eff_vs_flux} shows the muon detection efficiency as a function of the photon flux at various
high voltages. The efficiency decreases from 92$\%$ to 67$\%$ at 9.5 kV in a photon flux range from 18 
kHz/cm$^{2}$ to 403 kHz/cm$^{2}$. The conversion from the incident photon flux to the charged particle rate inside 
the gas gap requires detailed knowledge about the
interactions of photons with the materials between the source and the detector. We have
therefore adopted an approach based on the strip hit rate, which can be used as a proxy for the actual charged
particle rate for the purpose of the comparison to the expected rates in CBM. The strip hit rate was calculated as the average number of
hits per second per surface unit in the readout region of detector. Fig. \ref{hit_rate} shows a decrease of
efficiency with strip hit rate at 9.5 kV. The detector has been found to achieve a muon detection efficiency
higher than 90$\%$ and 83$\%$ at a photon-induced strip hit rate of $\sim$ 6 kHz/cm$^2$ and $\sim$ 10 kHz/cm$^2$ respectively.
\section{Discussions}
We have presented the beam test results for a low-resistivity bakelite RPC planned to be used in
the CBM experiment. A high voltage scan showed an efficiency plateau (at 95$\%$ efficiency) of 1 kV
above 9 kV. The time resolution of this RPC with MuCh-XYTER is measured to be 13 ns or less at 9.2 kV, which is in
good agreement with the requirements of CBM-MuCh. At a hit rate of $\sim$ 6 kHz/cm$^2$ and $\sim$ 10 kHz/cm$^2$, 
the efficiencies of the detector have been measured to be 90$\%$ and 83$\%$ at 9.5 kV. Given the
uncertainty in the calculated hit rate, further studies are required to make definitive conclusion on the rate capability
of the chamber. More studies are needed in order to have a fully efficient RPC at
rates such as those foreseen for the CBM stations. We list here possible developments that could lead to an improved rate capability.
Presently, at a threshold of 50 fC, we obtain an efficiency plateau starting at about 9 kV. Reduction of the noise in the system and thereby reducing the
threshold would allow us to operate the chamber with high efficiency at a lower gain, making use
of full bandwidth of the ASIC. It should be mentioned that the noise here refers to the system 
noise, not inherent detector noise. It could even be further improved by implementing the proper grounding scheme.
Thus a reduction of threshold will reduce the charge per hit ($\langle$ Q $\rangle$)
and hence improve the rate capability. Preliminary investigations suggest that with proper grounding scheme, the
threshold can be reduced to even lower values. In addition to this low-gain operation, we are also working on a
detector with thinner electrodes and smaller gas gap. Effort is also ongoing to optimize $\langle$ Q $\rangle$
by varying the percentages of iC$_4$H$_{10}$ and SF$_6$. We would, therefore, conclude that present results
show a path towards safe RPC operation both for the 3$^{rd}$ and 4$^{th}$ stations of CBM-MuCh.

\section{Acknowledgement}
We would like to thank the GIF++ test beam crew for providing us with excellent beam. The
Transition Radiation detector of CBM group, specially Philipp Kähler is acknowledged for providing
the space at GIF++ cave and their scintillator trigger logic. We would also like to thank the
entire CBM collaboration team for their enormous support in various directions. We thank Dr Rajesh
Ganai and Ms Ekata Nandy for their generous help in many ways.
\bibliographystyle{elsarticle-num} 
\bibliography{reference.bib}
\end{document}